\documentclass[twocolumn,prl,aps]{revtex4-1}
\usepackage{amsmath}
\usepackage{graphicx}
\usepackage{sistyle}

\newcommand{\erbium}{$\text{Er}^{3+}$}

\begin{document}
\title{Intrinsically stable light source at telecom wavelengths}

\author{Fernando Monteiro}
\author{Thiago Guerreiro}
\author{Bruno Sanguinetti}
\email{Bruno.Sanguinetti@unige.ch}
\author{Hugo Zbinden}
\address{Group of Applied Physics, University of Geneva, Chemin de Pinchat 22, CH-1211 Gen\`eve 4, Switzerland}

\begin{abstract}
We present a highly stable light source at telecom wavelengths, based on a short erbium doped fiber. The high stability arises from the high inversion of the \erbium ion population. This source is developed to work as a stable reference in radiometric applications and is useful in any application where high stability and/or a large bandwidth are necessary. The achieved long-term stability is \SI{10}{ppm}.
\end{abstract}
\maketitle

A great number of experiments ultimately rely on the measurement of an optical power. The precision of these experiments increases with the stability of its optical components, including the light source and the powermeter. 
In particular we are interested in specifying the performance and stability of components used in  fiber-based quantum communication systems.
Usually, such measurements are carried out using light from a coherent laser, calibrated against a standard powermeter and strongly attenuated. The systematic error incurred by this method of characterisation can typically reach 10\% when using commercial devices, and includes the uncertainty of the power meter's calibration, attenuator linearity, polarization and interference effects. When calibrating systems at the single-photon level, the power of the calibration laser is attenuated sometimes by more than ten orders of magnitude; small linearity errors in the attenuator can strongly contribute to systematic error.

The ideal calibration source is therefore broadband to eliminate interference effect, unpolarised, and emits at a power level which is high enough to be characterised  by a powermeter but low enough as not to require  strong attenuation when used to calibrate a single-photon detector. The source should also be intrinsically stable in the short and long terms.
Fiber sources \cite{Tran1996,High-Stability1997,Wang2011,Superfluorescent2013} have the advantage of single-mode operation, mechanical stability and a simple construction. They have been demonstrated to provide high spectral stability. However they are usually operated in the regime of Amplified Spontaneous Emission (ASE), which does not confer the power stability advantages of a purely spontaneous emission source.

Here we present an intrinsically stable light source based on spontaneous emission. This type of source can be used directly as part of an experiment, or can serve as a reference to characterize other optical elements.  This source is also well suited to being used in conjunction with fibre radiometers~\cite{Sanguinetti2010,Sanguinetti2012} to measure the absolute detection efficiencies of single-photon detectors.

First we describe the physical concepts behind the stability of such source and the experimental setup. We then examine the stability of each component taken individually and finally present our results and the long term stability of the source.

\begin{figure}[htbp]
\begin{center}
\includegraphics[width=8cm]{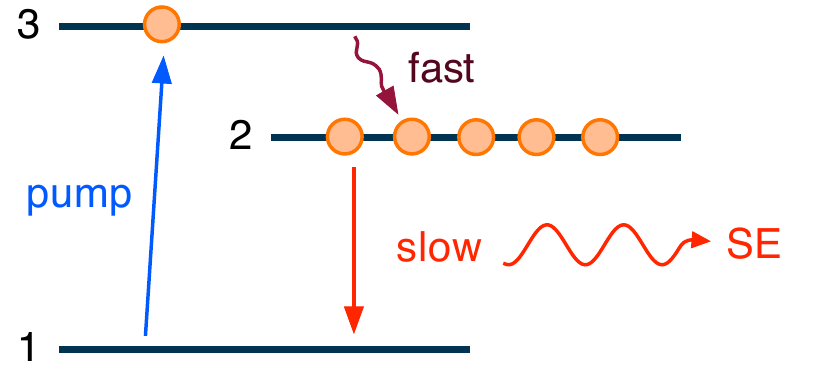}
\caption{Three-level model for \erbium spontaneous emission: A pump field brings atoms from the ground state $ \vert 1\rangle $ to an excited state $ \vert 3\rangle $ from which it rapidly decays to a metastable state $ \vert 2\rangle $. Atoms decay from level $ \vert 2\rangle $ to level $ \vert 1\rangle $ by spontaneous emission (SE).}
\label{default}
\end{center}
\end{figure}

\textit{Concept of the source:} The rate of spontaneous emission \cite{Scully1997} of an excited atom  depends on its structure and environment. In an erbium-doped fiber, the \erbium ions are physically fixed in their positions. In addition, the number $N$ of atoms in the fiber is also constant. For our purposes, the erbium structure can be modelled by a three level system~\cite{Peroni1990}, as show in figure \ref{default}. The electron in the ground state $ \vert 1\rangle $ absorbs the pump photon and goes to level $ \vert 3\rangle $. Then it rapidly decays into the metastable level $ \vert 2\rangle $ that decays to level $ \vert 1\rangle $ by emitting a photon at $ \sim $ \SI{1530}{nm}. Given that the pump power $P_\text{p}$ is much higher than the output power of the fiber (no pump depletion), the proportion $N^*/N$ of excited atom in level $ \vert 2\rangle $ with respect to level $ \vert 1\rangle $ is determined by the power $P_\text{p}$ of the pump laser~\cite{Peroni1990}
\begin{equation}
\label{eq:se_saturation}
N^*/N = P_\text{p} / (P_\text{p} + C)
\end{equation}
where $ C = \dfrac{h\nu}{\tau\sigma f} $; $ h $ is Planck's constant, $ \nu $ is the frequency of the pump laser, $ \sigma $ is the absorption cross section for the pump, $ \tau $ is the decay time from the metastable level $ \vert 2\rangle $ to the initial level $ \vert 1\rangle $ and $ f $ is the pump intensity profile at the fiber, such that $ 2\pi \int_{0}^{\infty} frdr = 1 $. This results in the spontaneous emission power $P_\text{SE}$ tending towards a constant as the pump power is increased.

\begin{figure}[h]
\centering
\includegraphics[width=8cm]{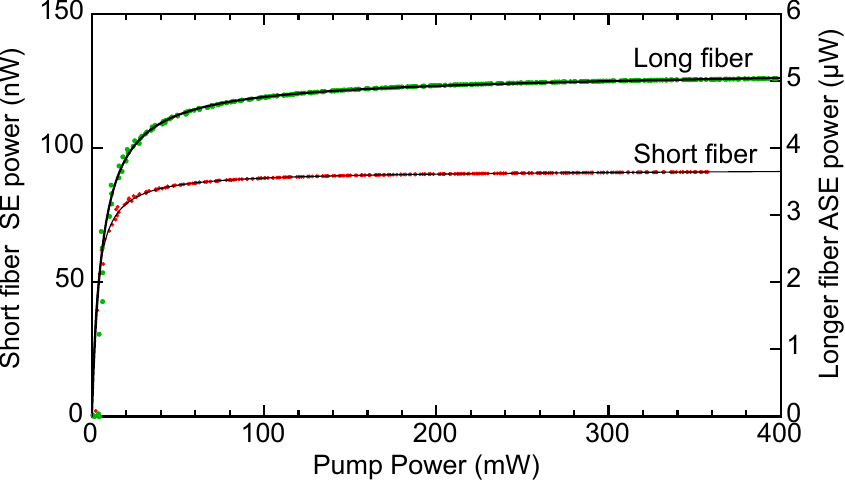}
\caption{\label{SE}
Saturation of the atomic medium: as the pump laser power $P_{P}$ is increased the spontaneous emission power $P_{SE}$ scale as $P_{SE}\propto P_\text{p} / (P_\text{p}+C)$ (fitted curves).}
\end{figure}

Figure \ref{SE} shows the experimental values for the spontaneous emission $P_\text{SE}$ versus $P_\text{p}$ for both a short and a long \erbium doped fiber. The $ P_{SE} $ scales as:
\begin{equation}
\label{eq: Power}
P_{SE} = \alpha N^*/N
\end{equation}
where $ \alpha $ is proportional to the length of the fiber, the number of \erbium per unit of length and also takes in account the coupling of the emitted light into the fiber and the amount of energy emitted per decay time.
The stability of spontaneous emission $P_{SE}$ is much higher than the stability in pump power $P_{P}$. We define a stability enhancement parameter $S$ as the relative change of pump power divided by the relative change in spontaneous output power: 
\begin{equation}
\label{eq: stability1}
S = \dfrac{\bigtriangleup P_{Pump}/P_{Pump}}{\bigtriangleup P_{SE}/P_{SE}} = \dfrac{\partial P_{P}}{\partial P_{SE}}\times \dfrac{P_{SE}}{P_{P}}
\end{equation}
and using equation (\ref{eq: Power}), we see that the stability depends linearly on the pump power:
\begin{equation}
\label{eq: stability2}
S(P_{P}) = 1 + \dfrac{\tau\sigma f }{h\nu} \times P_{P} \approx \dfrac{\tau\sigma f }{h\nu} \times P_{P}
\end{equation}
Using the cross section \SI{2.58E-25}{m^2} \cite{Desurvire1990}, the decay time \SI{10}{ms} \cite{Average1992,Desurvire1990} and assuming that $ f $ is flat over the core of the fiber and zero outside, together with the fiber specifications of Thorlabs Er30-4/125, we find $ S(\SI{0.35}{W}) = 134 \pm 18$. This is close to the value measured for the short fiber ($ \sim $ \SI{3}{mm}) $ S(\SI{0.35}{W}) = 99 \pm 10$ and around two times the value found for the long fiber ($ \sim $ \SI{10}{cm}) $ S(\SI{0.35}{W}) = 44 \pm 5$ of figure \ref{SE}. The discrepancy for the longer fiber can be explained by the fact that equations (\ref{eq:se_saturation}), (\ref{eq: Power}) and (\ref{eq: stability2}) only hold for spontaneous emission, while the output power of the longer fiber has a non negligible contribution from stimulated emission (amplified spontaneous emission, ASE). The amount of stimulated emission grows exponentially with fibre length, although for very short fibres ASE remains in the linear growth regime.

%The measured stability from the fitting also shows that the $ \sim $ \SI{3}{mm} fiber is two times more stable than the long fiber with respect to pump fluctuations.

\begin{figure}[htp]
\centering
\includegraphics[width=8cm]{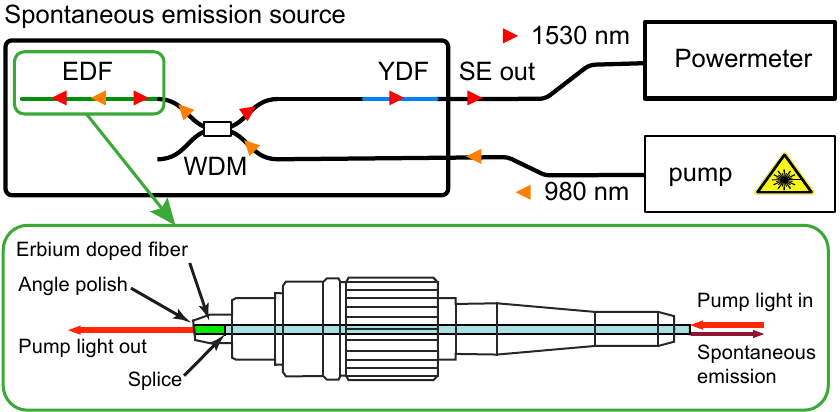}
\caption{Schematic view of the spontaneous emission source and $ \sim $ \SI{3}{mm} erbium doped fiber: a pump laser with a wavelength of \SI{980}{nm} is injected into an \erbium doped fiber (EDF) through a wavelength division multiplexer (WDM). The light which is spontaneously emitted in the backwards direction that couples into the fiber is directed to the output by the WDM. An Ytterbium-doped fiber is used to absorb any residual \SI{980}{nm} pump. The pump laser, the source and powermeter are temperature controlled.}
\label{setup}
\end{figure}

To achieve the required stability the source must be of simple construction, employing only spliced fiber components. 
An overview of the spontaneous emission source is given in Fig.~\ref{setup}. A short length of erbium-doped fiber is pumped by a diode laser at \SI{980}{nm}. Most of the pump light exits through the unconnected end of the EDF, which is angle polished in order to avoid back-reflections. The back-propagating spontaneous emission is extracted with a wavelength-division multiplexer (WDM) and an Ytterbium-doped fiber absorbs any residual \SI{980}{nm} light from the output. The source is temperature controlled to improve stability.

Despite the simplicity of the setup, critical attention to a number of points is necessary. Below we will describe each component in detail.

\textit{Erbium doped fiber:} We require the \erbium doped fiber to be short so that the emission is essentially spontaneous, the atomic medium is maximally inverted and there is no pump depletion, so that equation \ref{eq:se_saturation} holds. A short fiber has the additional benefit of being mechanically stable, having minimal polarisation effects, emitting a broad spectrum and  requiring little attenuation to achieve our desired power level.

In order to reliably integrate a very short section of \erbium doped fiber into our source, we employ the following procedure: first we splice a section of \erbium doped fiber (Thorlabs Er30-4/125) onto standard fiber. We then insert this fiber into a fiber connector, and apply epoxy so that only a millimetre-long section of doped fiber remains inside the connector. We then proceed to cleaving and angle-polishing in the standard way. 
A schematic view of the finished connector is shown in Figure \ref{setup}.

Most of the pump light exits the system from the angle-polished face of this connector and is absorbed by a metal cap which is thermally connected to a temperature-controlled metal plate. The back-reflections of the pump light are below \SI{-60}{dB}.

Using a low-coherence interferometer, we measured the coherence length to be \SI{20}{\mu m}, short enough to minimize interference effects. This corresponds to $1/\tau_c=\SI{1.4e13}{}$ modes per second. 
%The gain $G$ defined according to \cite{Sanguinetti2012} is
%\begin{equation}
%\mu_\text{out} = G\mu_{in} + G - 1
%\end{equation}
%where $\mu_\text{in}$, $\mu_\text{out}$ are the number of photons per mode on the input and output respectively. For a purely transparent medium $ G = 1 $ and for spontaneous emission (stimulated by the vacuum), $\mu_\text{in} = 0$, leading to:
%\begin{equation}
%\mu_\text{out} = G - 1
%\end{equation}
%The output power is of the order of \SI{100}{nW}, or \SI{6e11}{} photons per second. This corresponds to a $ \mu_\text{out} = 0.04 $ and gain of 1.04, indicating that we are working in the spontaneous emission regime.
At an output power of \SI{100}{nW}, the probability $p$ of a mode being populated by a photon is $0.04\ll 1$. This means that we are in the purely spontaneous emission regime; indeed amplified spontaneous emission would result in a strong probability of a mode being populated by at least 2 photons. The probability of this happening for a thermal state is $p^2$ and is negligible in our case.% confirming that the source operates in a purely spontaneous emission regime, as opposed to an amplified spontaneous emission regime.
 
\begin{figure}[htp]
\centering
\includegraphics[width=8cm]{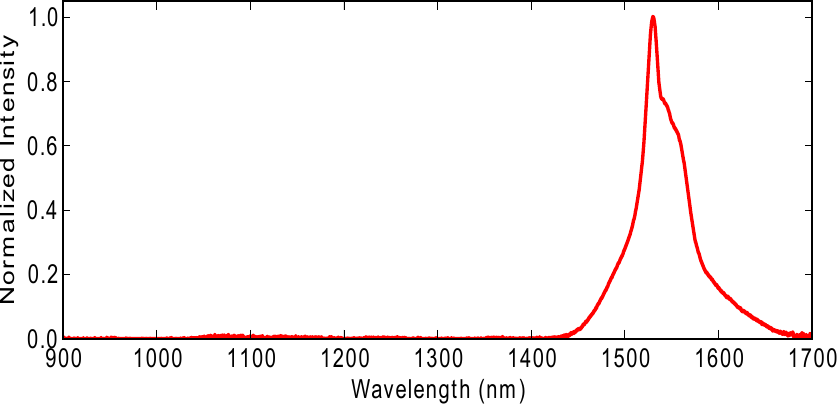}
\caption{Spectrum measured at the output of the stable source. 99 \% of the measured power is found between \SI{1400}{nm} and \SI{1700}{nm},and there is no peak at \SI{980}{nm} showing that the pump laser is filtered by the YDF.}
\label{spectrum}
\end{figure}

The measured coherence length is in agreement with the spectrum measured at the output of the source and shown in Figure~\ref{spectrum}. This spectrum peaks at \SI{1530}{nm} and presents no signature of the pump laser (\SI{980}{nm}), however, a small peak around $ \sim $ \SI{1050}{nm} to $ \sim $ \SI{1100}{nm} was observed, but its level is too close to the noise background to be identified. 

%Given the measured stability parameter $ S = 99 \pm 10 $ for this fiber, the normalized source standard deviation that we can achieve is given by the normalized pump standard deviation divided by 99.

The output power changes with the temperature of the splice and Erbium doped fiber by $ \bigtriangleup P_{SE} / P_{SE} = \SI{-71}{ppm / \degC}$.

\textit{WDM:} We used a WDM model AFW WDM-2-9815-L-1-L-0 to separate the pump from the output light. Changing the temperature of the WDM by 10$ ^{\circ} $C we observe a variation of output power of $ \bigtriangleup P_{SE} / P_{SE} = \SI{4.2}{ppm / \degC}$.

\textit{Ytterbium Doped fiber:} The measured transmission at the Ytterbium doped fiber (Thorlabs YB1200-4/125) for the pump wavelength is $ 2\times10^{-4} $ for powers below \SI{70}{\mu W}. The stability in temperature is $ \bigtriangleup P_{SE} / P_{SE} =\SI{-2.2}{ppm/\degC} $.

\textit{Pump Laser:} %To have a lower bound on the source standard deviation mentioned above, we put an effort in the stabilization of the pump laser.
The laser used is a JDS Uniphase model S30-7402-660. Its nominal operating power is \SI{720}{mW} at a current of \SI{1.3}{A}. We operate it at a constant current of \SI{900}{mA}, with an SRS LDC502 current controller. The output of this laser is very stable when measured at high powers on an integrating sphere. On a photodiode, however, interferometric and polarisation effects limit the measured stability. We did not notice a strong difference in stability between the constant current and constant power modes, although the constant power mode may perform better on the very long term (years), as the output power of pump lasers tends to decrease with usage.

To yet improve stability we control the temperature of the laser to $ 20.000 \pm 0.001 ^{\circ} C$. The normalized standard deviation of the measured power is $ 10^{-5} $ in one hour of measurement at $ \sim $ \SI{0.5}{W}. Taking into account the stability enhancement factor $S\approx 100$, this bounds the normalized standard deviation of our source in one hour of measurement to \SI{e-7}.

\textit{Powermeter:} When developing this source it became apparent that a standard powermeter would be the limiting factor to the precision and stability of the measurement. This limit comes both from the optics and electronics of standard powermeters. Optically, an integrating sphere with a fiber connector was found to perform well, but to require a large amount of optical power to generate enough photocurrent for the electronics to perform well. Standard powermeter diodes generate sufficient photocurrent but present some instability due to multiple reflections between the fiber connector face, photodiode surface and photodiode protection window. A trap detector \cite{Fox1991,Stock2000} would be a good solution, but the price for InGaAs trap detectors is prohibitive. We opted for a design which sends light on the diode (Hamamatsu G8605-12) at a slight angle, to avoid multiple reflections.
\begin{figure}[htbp]
\begin{center}
\includegraphics[width=8cm]{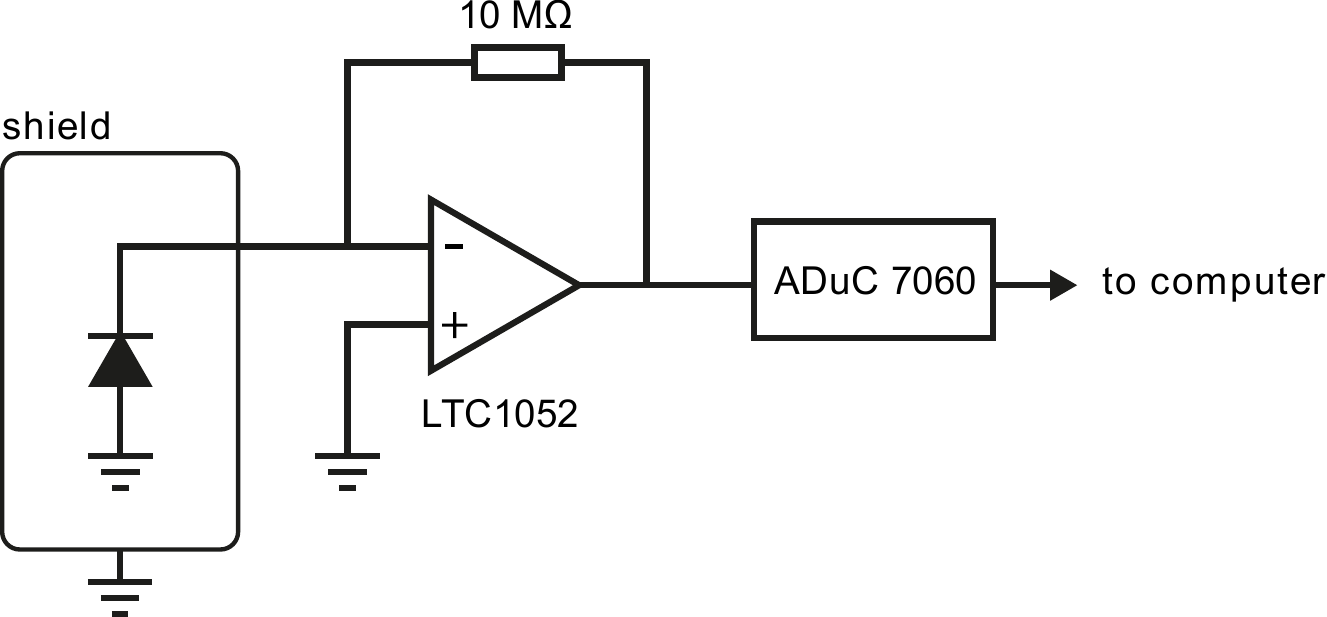}
\caption{Powermeter schematic: the photodiode is connected to a transimpedance amplifier. The output of this amplifier is acquired using a microconverter and set to a computer via an USB connection.}
\label{fig:powermeter}
\end{center}
\end{figure}

Electronic components often present temperature coefficients (tc) on the order of \SI{100}{ppm/\degC}. The most important electronics component is the transimpedence amplifier, which converts the photocurrent to a measurable voltage. Here a compromise must be made between low-noise devices and devices with a good long-term stability. The Linear Technology LTC1052 operational amplifier best suites this application, having a long term stability specified to better than \SI{100}{nV/\sqrt{month}} and and input noise current of \SI{0.6}{fA\sqrt{Hz}}. With the appropriate feedback resistor, this amplifier provides a trasimpedance gain of \SI{10}{M\ohm}, converting photocurrents of the order of \SI{100}{nA} to voltages of around \SI{1}{V}, suitable for acquisition by the \SI{24}{bit} digitiser of and Analog Devices ADuC7060 microconverter. A schematic representation of our powermeter is shown in figure \ref{fig:powermeter}. 

\begin{figure}[htp]
\centering
\includegraphics[width=9cm]{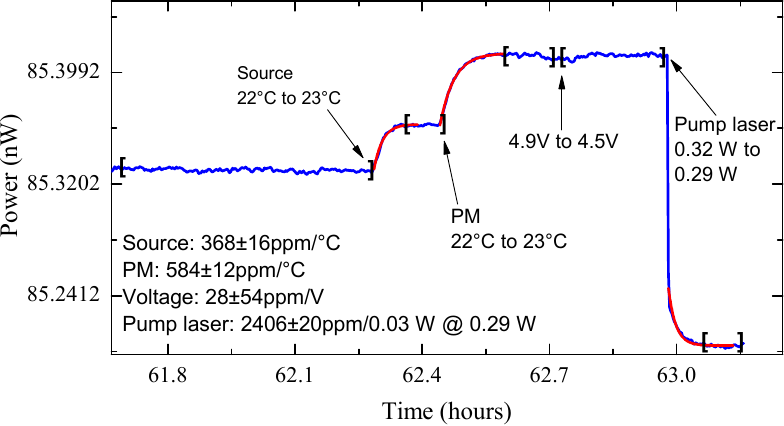}
\caption{Power variations by changing respectively the source temperature, the powermeter temperature, the voltage at the analogue to digital converter and the pump laser power.}
\label{timepower}
\end{figure}

We characterized how the source as a whole is affected by laser, powermeter and temperature variations. The temperature of the source was changed by \SI{1}{\degC}. This was followed by a temperature change of the powermeter of \SI{1}{\degC}. Then the power supply of the electronics was changed from $ 4.9 $ to \SI{4.5}{V} and the current of the pump laser from $ 889.5 $ to \SI{800}{mA}, that is $ 0.32 $ to \SI{0.29}{W} of pump power. The data is fit by $ P_{0} + A\times e^{-(t-t_{0})/\tau} $. Figure~\ref{timepower} shows the measured output power dependence, as measured with the powermeter, with respect to the temperature of source and powermeter, voltage of the analogue to digital converter and pump power. The typical time constant $ \tau $ of each fitted measured power is $ 72.5 \pm 3.7\ s $, $ 119.2 \pm 1.8\ s $, $ 62.3 \pm 4.4\ s $ for the source, powermeter and the change of the pump power respectively. Probably this last value is due to some temperature fluctuation caused by the decrease of pump power inside the components. This would explain why this value is close to the characteristic time when the temperature of the source changes. The assembled source and detector are more sensitive to temperature variations than expected from the characterisation of the individual components. However, note that the drifts of each element of the source cannot be added in quadrature, as they are all correlated to temperature changes.

\begin{figure}[h!]
\centering
\includegraphics[width=8cm]{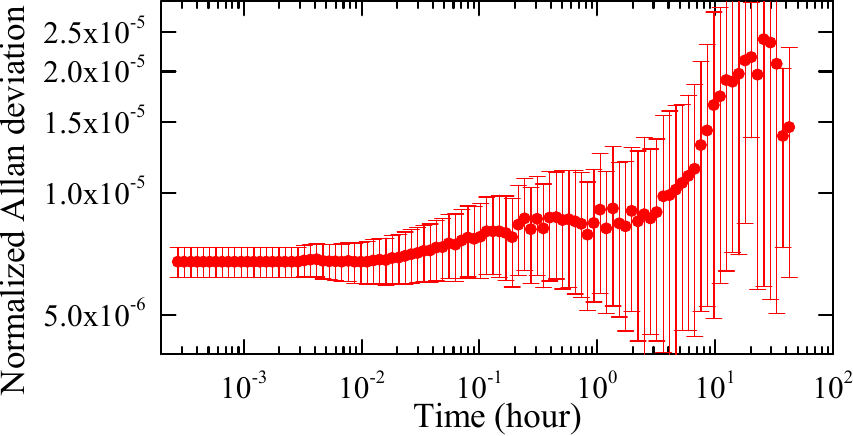}
\caption{Normalized Allan deviation over several days of measurement.}
\label{Allan}
\end{figure}

The normalized Allan deviation~\cite{Allan1966} from a four-days dataset is plotted in figure \ref{Allan}, showing a normalized Allan deviation of $ 8.4\times10^{-6} \pm 3.0\times10^{-6} $ in one hour and $ 1.5\times10^{-5} \pm 8.4\times10^{-6} $ in days of measurement. Figure \ref{timepower} shows that these Allan deviation values include contribution of both source and powermeter.

Assuming that the temperature fluctuations of powermeter and source are the same and using the data of figure \ref{timepower}, the calculated temperature fluctuation that explain such deviation is $ 1.7 \pm 0.6 \times 10^{-2}\ ^{\circ}C $ in one hour and $ 3.1 \pm 0.2 \times 10^{-2}\ ^{\circ}C $ in almost two days of measurement. These calculated values are in agreement with our expected temperature fluctuations.

In conclusion, our source and powermeter are ready for use just some minutes after being initialized and have a normalized standard deviation of $ 8.4\times10^{-6} \pm 3.0\times10^{-6} $ per hour of measurement. In principle is possible to achieve a normalized standard deviation of  $ 10^{-7} $ in one hour of measurement by improving temperature stability.

We thank Michael Afzelius for useful discussion, and are particularly grateful to Claudio Barreiro and Olivier Guinnard for their technical insights. We thank the Swiss NCCR QSIT for financial support.

\end{document}